# Persistent Enhancement of Exciton Diffusivity in CsPbBr$_3$ Nanocrystal Solids

Wenbi Shcherbakov-Wu[1], Seryio Saris[2,†], Thomas Sheehan[3], Narumi Nagaya Wong[3], Eric R. Powers[3], Franziska Krieg[4,5,††], Maksym V. Kovalenko[4,5], Adam P. Willard,[1] William A. Tisdale[*,3]

*1. Department of Chemistry, Massachusetts Institute of Technology, Cambridge, MA, United States*
*2. Laboratory of Nanochemistry for Energy (LNCE), Institute of Chemical Sciences and Engineering (ISIC), École Polytechnique Fédérale de Lausanne, CH-1950 Sion, Switzerland*
*3. Department of Chemical Engineering, Massachusetts Institute of Technology, Cambridge, MA, United States*
*4. Department of Chemistry and Applied Bioscience, ETH Zürich, Zürich, Switzerland*
*5. Laboratory for Thin Films and Photovoltaics and Laboratory for Transport at Nanoscale Interfaces, Empa – Swiss Federal Laboratories for Materials Science and Technology, Dübendorf, Switzerland*

*Correspondence: tisdale@mit.edu

†Current affiliation: Chair in Hybrid Nanosystems, Faculty of Physics, LMU Munich, Munich, Germany

††Current address: Avantama AG, Laubisrütistr. 50, 8712 Stafa Switzerland

**Abstract:**

In semiconductors, exciton or charge carrier diffusivity is typically described as an inherent material property. Here, we show that the transport of excitons (i.e., bound electron-hole pairs) in $CsPbBr_3$ perovskite nanocrystals (NCs) depends markedly on how recently those NCs were occupied by a previous exciton. Using fluence- and repetition-rate-dependent transient photoluminescence microscopy, we visualize the effect of excitation frequency on exciton transport in $CsPbBr_3$ NC solids. Surprisingly, we observe a striking dependence of the apparent exciton diffusivity on excitation laser power that does not arise from nonlinear exciton-exciton interactions nor from thermal heating of the sample. We interpret our observations with a model in which excitons cause NCs to undergo a transition to a metastable configuration that admits faster exciton transport by roughly an order of magnitude. This metastable configuration persists for ~microseconds at room temperature, and does not depend on the identity of surface ligands or presence of an oxide shell, suggesting that it is an intrinsic response of the perovskite lattice to electronic excitation. The exciton diffusivity observed here (>0.15 $cm^2$/s) is considerably higher than that observed in other NC systems on similar timescales, revealing unusually strong excitonic coupling in a NC material. The finding of a persistent enhancement in excitonic coupling between NCs may help explain other extraordinary photophysical behaviors observed in $CsPbBr_3$ NC arrays, such as superfluorescence. Additionally, faster exciton diffusivity under higher photoexcitation intensity is likely to provide practical insights for optoelectronic device engineering.

In semiconducting materials, the diffusion of charge carriers or excitons – which are bound electron-hole pairs – is central to the operation of electrical devices, generating energy in the form of electricity, light or heat. Recently, the development of time-resolved optical microscopy techniques has enabled direct visualization of electronic energy transport at the nanoscale.[1-11] In particular, experimental access to nonequilibrium regimes of exciton/carrier transport has revealed new insights into mesoscale dynamics. At early times, anomalous regimes of superdiffusive and/or hot-carrier transport may be observed.[12-13] At later times, the effect of energetic disorder manifests in subdiffusive transport phenomena, wherein the ensemble average diffusivity decreases over time.[14] Here, we report observation of a novel nonequilibrium transport modality in which the diffusivity of excitons in a nanocrystal array depends on *how recently those nanocrystals were previously in the excited state*.

Perovskite materials have attracted much attention over the past decade due to their potential applications in optoelectronic devices, such as photovoltaic cells and light-emitting diodes (LEDs).[15-22] Direct visualization of charge carrier and exciton diffusion has been investigated in bulk[6, 23-26] and two-dimensional (2D)[27-29] perovskites. $CsPbBr_3$ nanocrystals (NCs) are a particularly attractive perovskite morphology due to their bright and stable luminescence,[15, 30] quantum optical properties,[31-33] and evidence for strong excitonic coupling in NC arrays.[34-36] Indeed, initial reports suggest highly mobile excitons within $CsPbBr_3$ NC solids.[7, 37-38]

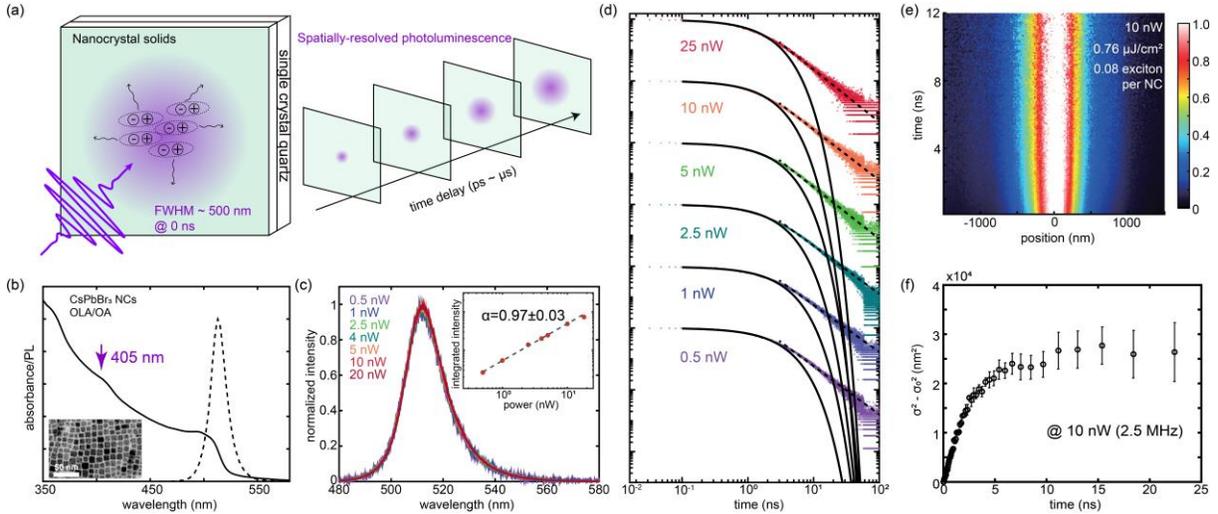

**Fig. 1.** *Transient photoluminescence microscopy (TPLM) of CsPbBr$_3$ NC solids.* (a) Schematic showing a near diffraction-limited laser pulse generating a population of excitons, which subsequently diffuse within the film, leading to spatial broadening of the photoluminescence signal over time (laser excitation spot not to scale). (b) Normalized absorbance (solid line) and photoluminescence (dashed line) spectra of the CsPbBr$_3$ (OLA-OA) NCs in toluene. Arrow indicates the laser excitation wavelength for all of the TPLM measurements in this study. Inset shows a transmission electron micrograph (TEM) image of the CsPbBr$_3$ NC sample. (c) Normalized power-dependent emission spectra showing no peak shift as a function of excitation laser power (0.5~20 nW, 2.5 MHz, 500 nm diameter laser spot). Inset shows the integrated photoluminescence intensity as a function of laser excitation power. Dashed line is a linear fit of the data with fitted parameters labeled above. (d) Power-dependent transient photoluminescence plotted on log-log scale (5 MHz repetition rate). The solid lines are single-exponential fits to the first 3 ns, and dashed lines are power law fits to 3~20 ns. (e) Sample normalized TPLM color map plotted as a function of spatial position (x-axis) and time (y-axis) at 10 nW (2.5 MHz repetition rate). (f) Mean square displacement as a function of time extracted from the data shown in (e). See Table S1-2 for detailed measurement parameters.

To characterize exciton dynamics in CsPbBr$_3$ NC solids, we employed transient photoluminescence microscopy (TPLM) to track radiative recombination of photogenerated excitons with both temporal and spatial resolution, as illustrated in **Fig. 1a & S1**. Briefly, a variable repetition rate pulsed laser (405 nm, ~50 ps, FWHM 500 nm, 0.5-40 MHz) generated an initial population of excitons in a near-diffraction-limited spot. Epifluorescence was collected by a microscope objective lens and magnified by ~500x using a telescope, then an avalanche photodiode (APD, 50 µm × 50 µm active area) was raster scanned across the magnified imaging plane. Transient photoluminescence data were collected at each spatial position, allowing the time-dependent spatial distribution to be re-constructed and analyzed. The overall temporal resolution (~80 ps) is limited by the excitation laser pulse width and the APD response time. As a superresolution optical technique, the spatial resolution is ultimately limited by the total photon counts and other ancillary factors impacting the signal-to-noise ratio of the measurement, meaning that exciton diffusion lengths much smaller than the focused laser beam waist can be reliably determined for bright emitters.[2] A diagram of the instrument and further experimental details (Supplementary Note 3, **Fig. S1-2**) can be found in the Supporting Information.

We first investigated $CsPbBr_3$ NCs capped with a mixture of oleylamine and oleic acid (OLA/OA) surface ligands, which were spun-cast into thin solid films supported on quartz glass substrates. The quasi-cubic $CsPbBr_3$ NCs were 8.3 nm in size, and the solid film exhibited a photoluminescence quantum yield of 68%. The NC film absorption and emission spectra are shown in **Fig. 1b**. The transmission electron micrograph (TEM) image shows the quasi-cubic shape of the NCs (**Fig. 1b inset**). The thin film samples were 30-40 nm in thickness and relatively uniform, as confirmed by atomic force microscopy (AFM) (**Fig. S4**).

To ensure that TPLM experiments were conducted in the linear regime, we performed power-dependent photoluminescence spectroscopy to verify the absence of nonlinear multi-exciton interactions. **Fig. 1c** shows normalized photoluminescence spectra collected with nominal laser power ranging from 0.5 to 20 nW at a constant repetition rate of 2.5 MHz focused to a ~500 nm diameter laser spot, corresponding to 0.004 to 0.16 photons absorbed per NC per laser pulse – see **Table S1**. Across this power regime, no peak shifting or broadening is observed; more importantly, there is a linear relationship between integrated photoluminescence intensity and excitation laser power, indicating that additional non-radiative multi-exciton (i.e. Auger) processes are not introduced at higher power. The same linear trend was also observed when excitation laser power was increased under constant laser fluence and varying repetition rate (**Fig. S5**).

As observed by others,[15] the transient photoluminescence decay of CsPbBr$_3$ NCs exhibits both prompt and delayed emission characteristics (**Fig. 1d, Fig. S6**). The prompt emission, covering the first ~0-3 ns following photoexcitation, is well fit by a single-exponential decay curve (solid black lines on the log-log plot in **Fig. 1d**), while the delayed emission (≥3 ns) can be described by a power-law (dashed line). While the origin of these complex emission dynamics is still debated, single-NC studies suggest that the prompt emission results from direct exciton recombination, whereas the delayed power-law emission arises from excitons that have undergone multiple trapping-detrapping processes.[39] With increasing laser power, the prompt emission lifetime slightly decreased from ~2.3 ns to ~1.5 ns, while the power-law exponent for the delayed emission increased from ~1.5 to ~1.7 (**Fig. S6**).

The results of one TPLM measurement are shown in **Fig. 1e**. These data were collected with a nominal laser power of 10 nW at 2.5 MHz repetition rate and 500 nm laser spot size, corresponding to a laser pulse fluence of ~0.76 µJ/cm$^2$, or 0.08 photogenerated excitons per NC per pulse (see Supplementary Note 6). To quantitatively extract exciton diffusivity from the TPLM data shown in **Fig. 1e**, the instantaneous spatial profile was fitted to a Gaussian shape at each delay time and the mean square displacement (MSD), equal to the change in variance of the Gaussian distribution $\sigma(t)^2 - \sigma_0^2$, was extracted (Supplementary Notes 3, **Fig. S2**).[2] The diffusivity of the exciton population is then extracted from the slope of the MSD curve. For normal diffusion, the MSD follows a linear relationship with time:

$$\sigma^2(t) - \sigma^2(t=0) = 2Dt \qquad (1)$$

where $D$ is the diffusivity.

The MSD curve generated from the TPLM data in **Fig. 1e** is shown in **Fig. 1f**. Interestingly, the MSD grows linearly with time during the first ~3 ns then becomes sublinear afterwards, matching the transition from prompt to delayed emission in the transient photoluminescence data (**Fig. 1d**). The MSD behavior is consistent with the transient photoluminescence decay interpretation: at early times excitons diffuse freely within the NC solid while spontaneously undergoing radiative recombination; at later times, exciton trapping begins to dominate the spatiotemporal dynamics leading to a dramatic decrease in exciton diffusivity. For the remainder of this study, we focus exclusively on the early-time dynamics (<3~4 ns), corresponding to free exciton diffusion.

TPLM experiments were performed while independently varying the laser repetition rate, laser pulse fluence (proportional to the excitation density, $\langle N \rangle$; see Supplementary Note 6), and, accordingly, the time-averaged laser power (**Fig. 2**). We observed that increasing either the fluence or repetition rate while holding the other variable constant always led to a greater measured value of the exciton diffusivity (**Fig. 2a-d**). Surprisingly, if the fluence and repetition rate were varied simultaneously such that the time-average power remained constant, the measured value of the exciton diffusivity did not change (**Fig. 2e**, **Fig. S3**). TPLM experiments performed under a variety of laser excitation conditions are aggregated and plotted together in **Fig. 2f-h**. While there is no correlation of diffusivity with either laser pulse fluence or laser repetition rate alone (**Fig. 2g,h**), there is a strong positive correlation between diffusivity and time-averaged laser power (**Fig. 2f**).

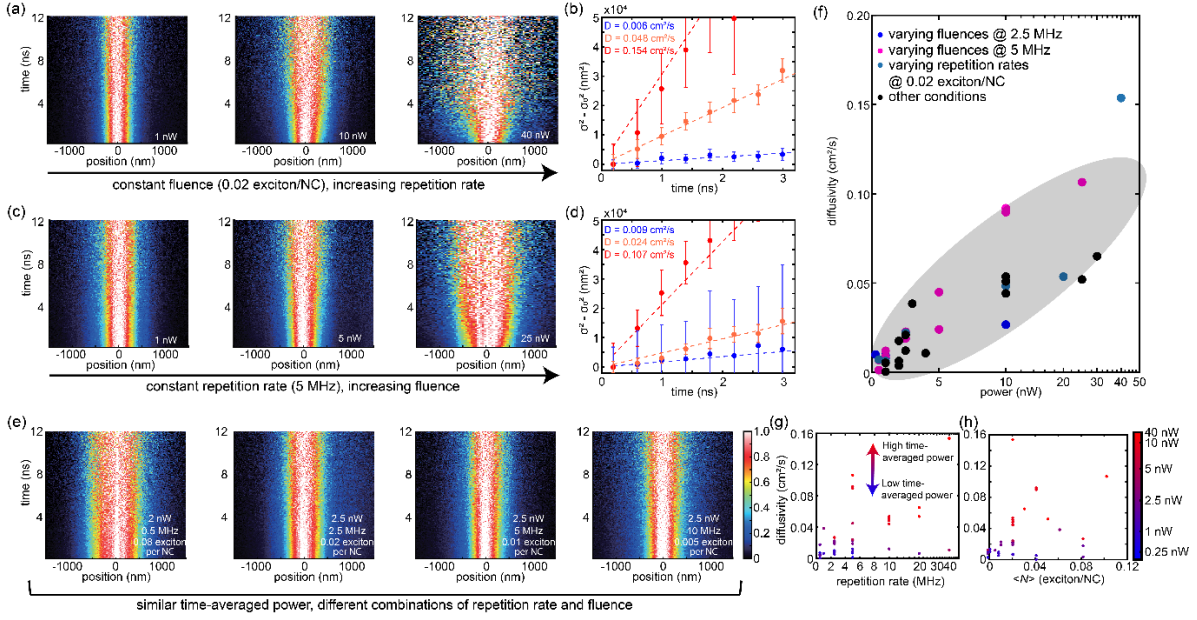

**Fig. 2.** *Dependence of measured exciton diffusivity on laser pulse fluence, repetition rate, and time-averaged power.* (a,b) TPLM data with varying repetition rate, while holding pulse fluence constant at $\langle N \rangle = 0.02$ absorbed photons per NC per laser pulse. (c,d) TPLM data with varying pulse fluence, while holding repetition rate constant (5 MHz). (e) TPLM data while varying pulse fluence and repetition rate together to achieve constant time-averaged power. (f) Exciton diffusivity measured under varying experimental conditions plotted *vs* time-averaged power. The grey oval is a guide to the eye. The x-axis is in linear scale when power is less than 10 nW and logarithm scale when power is above 10 nW. (g,h) Exciton diffusivity measured under varying experimental conditions plotted *vs* laser repetition rate or $\langle N \rangle$. Data points are color-coded according to time-average power, indicated to the right.

The observation of power-dependent diffusivity in the low-excitation-density regime is surprising. Typically, exciton transport in semiconductor NC solids is understood in the framework of incoherent hopping mediated through dipole-dipole interactions, *i.e.* Förster Resonant Energy Transfer (FRET).[7, 14, 40-42] In this picture, excitons act effectively as point dipoles that undergo discrete stochastic hopping transitions to neighboring NCs; varying the excitation laser power is expected to have no effect on the rate at which excitons move. We note that, at sufficiently high excitation density, $\langle N \rangle$, exciton-exciton interactions will introduce additional nonlinear recombination pathways that can interfere with the interpretation of time-resolved microscopy measurements;[2-3] however, no dependence of the measured diffusivity on laser pulse fluence alone was observed (**Fig. 2h**). The correlation between exciton diffusivity and time-averaged laser power (**Fig. 2f**) – but *not* repetition rate or laser pulse fluence alone – cannot be explained within the traditional exciton random walk model. Consequently, we explored multiple experimental and materials factors that might contribute to this anomalous observation, including 1) sample degradation, 2) instrumentation/data analysis artifacts, 3) laser heating, and 4) surface effects.

We first sought to verify that the power-dependent trend shown in **Fig. 2f** is repeatable and reversible. A series of TPLM measurements was performed at the same spot on the sample while non-monotonically varying laser power (**Fig. 3a**). We established a low-power baseline diffusivity by starting the measurement at 1.5 nW, and then raised the power to 6 nW. The TPLM scan was repeated again at 6 nW, then the power was lowered back to 1.5 nW, confirming that the low-power diffusivity value could be recovered after the same sample spot was exposed to higher laser power. The last two runs further confirmed the general reproducibility of the measurement, while also revealing some of the scatter in the data evident in **Fig. 2f**.

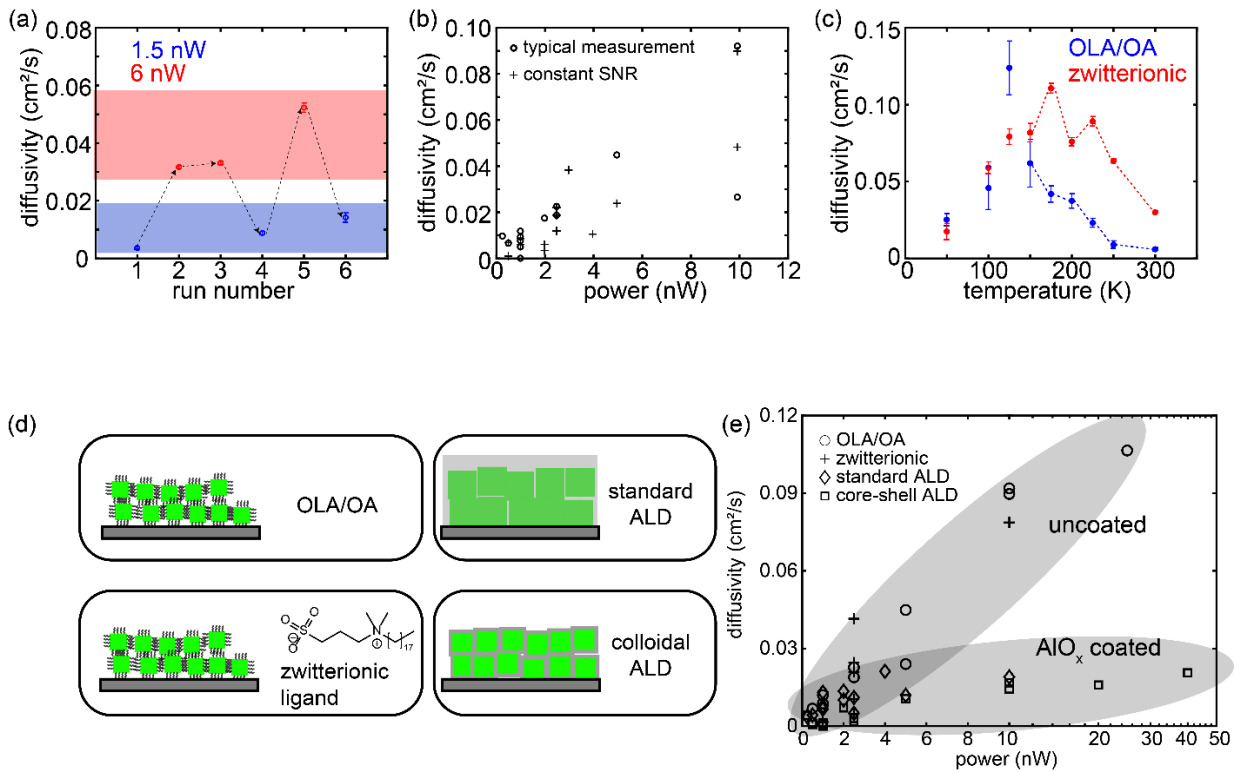

**Fig. 3.** *Measurement repeatability, and effect of temperature, signal-to-noise ratio, and surface chemistry.* (a) Demonstration that power-dependent diffusivity is a repeatable and reversible observation. TPLM was performed at the same sample spot while varying the excitation power non-monotonically. (b) Invariance of the power-dependent trend with measurement signal-to-noise ratio. Open circles correspond to the typical TPLM measurement, in which signal count rate varies naturally with the laser power used. Crosses correspond to a power series in which variable neutral density filters were placed in the signal path to achieve a constant signal-to-noise ratio across all powers. (c) Temperature-dependent exciton diffusivity of $CsPbBr_3$ NCs capped with different ligands, as measured at 1 nW and 2.5 MHz. (d) Illustration of four different surface treatments investigated. Top left shows the $CsPbBr_3$ NCs capped with oleylammonium and oleate ligands (OLA/OA). Bottom left shows the NC samples capped with zwitterionic ligands, 3-(N,N-dimethyloctadecylammonio)-propanesulfonate. Top right shows the NC film coated with $AlO_x$ using standard atomic layer deposition (ALD). Bottom right shows the NC sample coated with AlOx using a colloidal ALD growth. (e) Correlation of exciton diffusivity with time-averaged laser power for $CsPbBr_3$ NC solids with different surface treatments. The x-axis is in linear scale when power is less than 10 nW and logarithm scale when power is above 10 nW. The grey ovals are a guide to the eye, separately grouping the uncoated and $AlO_x$-coated samples.

Next, we investigated whether varying signal-to-noise ratio of the TPLM measurement at different laser power could lead to systematic errors in data analysis. In **Fig. 3b** we compare a typical TPLM measurement series (open circles) to one in which variable neutral density filters were placed in the detection pathway to achieve a constant signal-to-noise ratio across all measurements (crosses). The two data sets overlap completely (albeit with some scatter in the data), demonstrating that differences in signal intensity or APD count rate are not responsible for the power-dependent trend. Moreover, we tested our data analysis procedure on simulated data with varying shot noise and background noise and found that diffusivities below 0.01 cm$^2$/s could be reliably extracted from experimental data of the quality presented here (Supplementary Note 7).

Perhaps the biggest concern in any power-dependent spectroscopy trend is possible effects of laser heating. When a laser pulse is absorbed by the sample, heat is generated as the excess photon energy is dissipated over varying timescales. Typically, carrier-carrier scattering first leads to formation of a Boltzmann distribution over the electronic degrees of freedom within ~100 fs,[43] followed by hot carrier relaxation to the band edge *via* phonon emission on a picosecond timescale.[44] As time-averaged power rises, more heat is generated, increasing the lattice temperature. The temperature rise is ultimately limited by thermal transport away from the laser excitation spot, a process whose timescale depends on the geometry of the measurement and the thermal conductivity of the sample. Thermal transport simulations of our TPLM experiment showed that, under typical laser fluence used here (~0.1 µJ/cm$^2$), each absorbed laser pulse increases the local sample temperature by less than 0.01 K – with the majority of that heat dissipating within the first few nanoseconds following photoexcitation (Supplementary Note 8). These findings are consistent with experimental results obtained using time-resolved X-ray diffraction by Kirschner *et al.*, who found that excitation fluences on the order of a few mJ/cm$^2$ (~10,000x larger than that used in our TPLM experiments) were required to increase the temperature of CsPbBr$_3$ NCs by ~100 K, and that the heat fully dissipated within ~10 ns.[45]

To further investigate the potential consequences of sample heating, we directly measured the exciton diffusivity as a function of sample temperature (**Fig. 3c**). Temperature-dependent TPLM measurements were performed inside a closed-cycle liquid helium cryostat under vacuum (Montana Instruments Cryostation). Two CsPbBr$_3$ NC samples having similar size – but different surface chemistry – were investigated. One sample was terminated with a mixture of oleylammonium and oleate ligands (OLA/OA), while the other sample was terminated with a zwitterionic ligand, 3-(*N,N*-dimethyloctadecylammonio)propanesulfonate.[46] In both samples, the diffusivity exhibited a similar non-monotonic trend – first increasing as the temperature decreased below room temperature, then eventually decreasing again as the sample was further cooled below ~150 K. The full temperature-dependent behavior is the subject of ongoing investigation, but the salient observation here is the trend of *decreasing* diffusivity with *increasing* sample temperature near room temperature. If laser heating were responsible for the power-dependent trend shown in **Fig. 2f** and **Fig. 3e**, we would expect the opposite behavior. Consequently, we conclude that laser heating is not responsible for the anomalous power-dependent trend in the data.

Unusual behavior in semiconductor nanocrystals is sometimes associated with their unique surface properties. Next, we examined whether varying surface treatments led to the same power-dependent phenomenon. Four different types of samples were investigated: 1) the OLA/OA-capped NCs shown in **Fig. 1**, 2) NCs synthesized with a zwitterionic ligand, [3-(N,N-dimethyloctadecylammonio)-propanesulfonate], 3) spun-cast NC solids coated with $AlO_x$ using standard atomic layer deposition (ALD),[47] and 4) NCs coated with an $AlO_x$ shell grown *via* colloidal ALD prior to spin-casting,[48] as illustrated in **Fig. 3d**. Samples #1, 3, and 4 were prepared at EPFL (Lausanne), while sample #2 was prepared at ETH (Zurich), before being shipped to MIT for transient microscopy measurements. Optical characterization of all samples is reported in **Fig. S9**, and details of the sample preparation methods used are included in the Supplementary Information.

In **Fig. 3e** we compare the results of TPLM measurements made on the four samples with different surface treatments. All four samples show the same monotonic trend of increasing exciton diffusivity with increasing laser power. Similar to the OLA/OA samples shown in **Fig. 2**, there was no correlation between exciton diffusivity and laser pulse fluence or repetition rate alone (**Fig. S10**). Importantly, the two samples with $AlO_x$ coatings exhibited lower absolute diffusivity than the uncoated NCs, which is consistent with previous studies demonstrating slower exciton transfer rates between NCs with thicker surface shells.[14] Moreover, when the colloidal $AlO_x$ shell thickness increased from 4 layers to 8 or 12 layers, the diffusivity became too small to measure (**Fig. S11**). These findings build confidence in the measurement technique and its connection to the physics of exciton transport. Meanwhile, the repeated observation of power-dependent diffusivity in multiple CsPbBr$_3$ NC samples synthesized by different labs and terminated with different chemistries suggests that this unexpected correlation is intrinsic to the CsPbBr$_3$ perovskite lattice – and not derived from surface-related phenomena.

To understand the observation of power-dependent diffusivity, we consider the microscopic meaning of time-averaged laser power in the context of this experiment. In the low excitation density regime (i.e. <*N*> less than 1), time-averaged power informs on the frequency of NC photoexcitation events; specifically, the inverse of time-averaged power is proportional to the average waiting time between NC excitation events. In **Fig. 4a**, we plot the measured exciton diffusivity *vs.* time between NC excitations (proportional to 1/*power*). When plotted in this way, the data exhibit a characteristic exponential relaxation curve, with an extracted relaxation time constant of ~6 μs. As the waiting time between individual excitation events becomes longer, the observed ensemble exciton diffusivity decreases. The limiting diffusivity (as time between excitation events tends toward infinity) is ~0.01 cm$^2$/s, while the "enhanced" diffusivity measured under frequent photoexcitation conditions is up to 15x larger. The observation that exciton diffusivity changes as the exciton generation frequency on any given NC changes is surprising. Significantly, the relaxation time constant – the characteristic time it takes for a NC to relax to the low-diffusivity state – is on the microsecond timescale, which is 2-3 orders of magnitude longer than the exciton lifetime (**Fig. 1**). This striking observation suggests that CsPbBr$_3$ NCs retain some persistent memory of previous exciton occupation events.

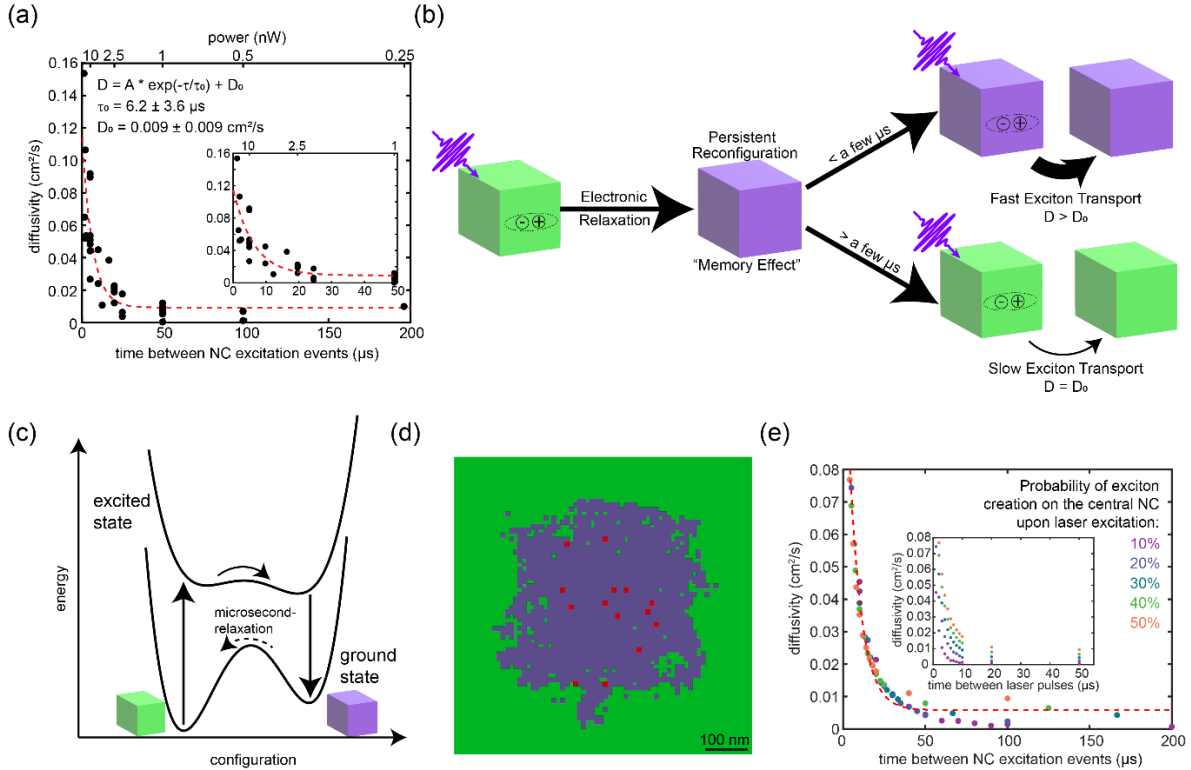

**Fig. 4.** *Persistent enhancement of exciton diffusivity.* (a) Experimentally measured diffusivity relaxation curve. Exciton diffusivity of CsPbBr$_3$ NCs with OLA/OA ligands as a function of time between NC excitation events. Mirror *x*-axis shows the corresponding time-averaged power. The red dashed line is a single exponential fit to the data, and the best fit parameters are annotated within the figure. Inset is a magnification of the early-time data points. (b) Schematic illustration of the excitation memory effect leading to persistent enhancement of exciton diffusivity. (c) Potential energy surface description of the phenomenon illustrated in panel (b). Black curves indicate the electronic excited state and electronic ground state of the NCs. (d) Snapshot of a kinetic Monte Carlo (KMC) simulation of exciton transport in a 2D NC array, which includes excitation memory effects. NCs in the relaxed state are shown in green and NCs in the metastable state are shown in purple. Red spots indicate current location of excitons within the time-dependent simulation. (e) KMC simulation results plotted as a function of time between NC excitation events, showing consistency with the experimentally measured phenomenon. Different colors represent different simulated excitation fluences. Inset shows simulated diffusivity as a function of time between laser pulses (reciprocal of laser repetition rate). The red dashed line is a single exponential fit to the data.

The exponential relaxation behavior shown in **Fig. 4a** suggests that NCs can adopt two different states: a stable "slow state" associated with the low-diffusivity exciton transport regime, and a metastable "fast state" that permits dramatically increased exciton hopping rates. These states are illustrated schematically in **Fig. 4b,c**. To test our understanding, we built a phenomenological model of exciton transport that includes persistent excitation memory effects and studied model dynamics using kinetic Monte Carlo (KMC) simulations (**Fig. 4d,e**). In the model, the NC solid is represented as a two-dimensional square lattice - each lattice site representing a single NC – with lattice spacing, $l \approx 10 nm$, roughly corresponding to the neareast neighbor spacing.

In the model, NCs can adopt either a *slow* or *fast* state, depicted in **Fig. 4d** as green or purple squares, respectively. Excitons, depicted in red in **Fig. 4d**, occupy individual NCs and can hop between nearest neighbor NCs with a rate $k_{\text{hop}}$ that depends on the state of the neighboring NCs. Excitons hop to *slow* or *fast* state NCs with rates $k_{\text{hop}}^{(\text{slow})} = 0.006 \text{ps}^{-1}$ or $k_{\text{hop}}^{(\text{fast})} = 0.7 \text{ps}^{-1}$, respectively. The rates were empirically chosen to match the experiments. Excitons decay stochastically with a probability determined by their lifetime, $\tau_{\text{ex}} = 2.5 \text{ns}$. We model the creation of excitons via laser pulse by randomly exciting sites with a Gaussian spatial distribution whose width, $\sigma_{\text{pulse}}$, and amplitude, $\alpha_{\text{pulse}}$, are selected to match experimental estimates. Subsequent pulses create new excitons at regular intervals whose spacing in time far exceeds the exciton lifetime, so that no excitons from previous pulses remain.

The state of a NC (*i.e.*, *slow* or *fast*) depends on the history of its exciton occupancy. Upon excitation – *via* laser pulse or hopping – an NC immediately transitions into the *fast* state and can remain in this state even if the exciton decays or hops away. Unoccupied *fast* state NCs decay stochastically back to the *slow* state with a rate of $k_{\text{relax}} = (8.1 \, \mu s)^{-1}$. Due to the difference between exciton (~ns) and *fast* state (~µs) lifetimes, all NC relaxation is assumed to occur in the time between laser pulses, after all excitons have decayed. **Fig. 4d** illustrates a snapshot of the simulation a few nanoseconds after one of the laser pulses.

We compute exciton diffusivity from the simulation in a manner analogous to the experiment. Specifically, we compute the mean radial exciton density as a function of time following the most recent laser pulse, $\rho_{\text{ex}}(r, t)$, assuming the center of the laser pulse is located at the origin. We extract the time-dependent width of this distribution, $\sigma^2(t) = \langle r^2(t) \rangle$, where the angle brackets imply an average over $\rho_{\text{ex}}$, and a simulated diffusivity as $D_{\text{sim}} = (\sigma^2(t) - \sigma^2(0))/2t$, evaluated at $t = 1\text{ns}$. **Fig. 4e** plots the simulated values of $D_{\text{sim}}$ for a range of simulations with differing pulse frequency and pulse intensity. We find that $D_{\text{sim}}$ is primarily a function of the average time between NC photoexcitation as evaluated at the center of the Gaussian laser pulse. This result is in near quantitative agreement with experimental observation.

Exciton/charge carrier dynamics in NCs on microsecond timescales has often been discussed in the framework of photocharging, wherein an individual charge carrier (electron or hole) can remain on a NC for a few microseconds before full relaxation.[49-52] Photocharging in CsPbBr$_3$ NCs has been observed at higher laser pulse fluences at cryogenic temperatures,[39] and is typically associated with electron (or hole) transfer to the NC surface. To investigate this possibility, we compared the diffusivity relaxation curves for OLA/OA-capped NCs (shown in **Fig. 4a**) to the colloidal AlO$_x$-coated CsPbBr$_3$ NCs, shown in **Fig. S12**. Despite the presence of the insulating oxide layer on the AlO$_x$-coated NC surface, the relaxation time constant was 10.8 μs – only slightly larger than the 6 μs value measured for the uncoated NCs. Moreover, photocharging is typically a nonlinear effect, requiring interaction of one exciton with another exciton or photon, which we have shown is not the case in this study (**Fig. 2h**). Finally, it is not clear why a charged NC should exhibit dramatically enhanced excitonic coupling to its neighbors.

A more provocative explanation for the excitation memory effect illustrated in **Fig. 4b,c** is the presence of a lattice polarization that persists long after exciton recombination. The existence of polarons – lattice deformations coupled to electronic excitations – is frequently invoked to explain experimental observations in halide perovskites.[53-58] Particularly intriguing is the possibility of microscopic ferroelectric domains, which could act to collapse the oscillator strength along a particular NC axis. In this picture, the transition back to the relaxed/disordered state requires thermal energy from the surrounding which results in a slow relaxation time, as is shown in **Fig. 4b,c**. However, the presence of ferroelectric-like behavior in halide perovskites is debated.[59]

Regardless of the microscopic mechanism, the phenomenological observation of a persistent enhancement in excitonic coupling in CsPbBr$_3$ NC solids is significant. In the low power limit, exciton diffusivity was measured to be around 0.01 cm$^2$/s, which is slightly higher than the corresponding value in CdSe QD solids.[60-64] However, under the highest laser powers used in this study, the diffusivity reached 0.15 cm$^2$/s (we note that Penzo *et al.* reported a diffusivity of 0.5 cm$^2$/s in similar CsPbBr$_3$ NCs).[7] For reference, carrier diffusivity in bulk halide perovskite single crystals at room temperature has been estimated between 0.3-0.5 cm$^2$/s, while diffusivity in perovskite thin films is typically an order of magnitude smaller, ~0.01-0.05 cm$^2$/s.[25-26, 65-66] The observation of an exciton diffusivity in CsPbBr$_3$ NC solids that exceeds the carrier diffusivity in perovskite thin films – despite the presence of long-chain organic ligands separating individual NCs – is striking. Our findings may also inform other observations of strong excitonic coupling in CsPbBr$_3$ NCs. In particular, superfluorescence – the collective emission of excitons coherently coupled across multiple NCs – has been observed in both CsPbBr$_3$ NC superlattices and binary NC superlattices.[34-35] Different from excitonic coherences sometimes observed in molecular systems,[67] the term superfluorescence implies the presence of strong excitonic coupling in the excited state – but not the ground state. Temporary transition to a transient lattice configuration within the excited state manifold in which excitonic coupling is enhanced could help explain this behavior. Finally, the observation of power-dependent diffusivity has implications for the design of high-brightness LEDs[21, 68-75] and lasers[76-78] featuring $CsPbBr_3$ NCs.

## Associated content

**Supporting Information**
>Detailed experimental methods, sample characterizations, and supplementary notes and discussion.

## Author information
**Corresponding Authors:**
William A. Tisdale: tisdale@mit.edu

**ORCID**
Wenbi Shcherbakov-Wu: 0000-0001-8898-1074
Seryio Saris: 0000-0001-8498-5852
Thomas Sheehan: 0000-0002-9540-3915
Narumi Nagaya Wong: 0000-0002-7192-3502
Eric R. Powers: 0000-0003-1342-5801
Raffaella Buonsanti: 0000-0002-6592-1869
Franziska Krieg: 0000-0002-0370-1318
Maksym V. Kovalenko: 0000-0002-6396-8938
Adam P. Willard: 0000-0002-0934-4737
William A. Tisdale: 0000-0002-6615-5342

**Notes:**
The authors declare no competing financial interest.

## Acknowledgement
We thank Dr. Raffaella Buonsanti for providing the AlOx-coated $CsPbBr_3$ NC samples and helpful discussion. We thank Dr. Anna Loiudice for preparing $CsPbBr_3$ (OLA/OA) NC solid samples. We thank Dr. Dmitry Dirin for preparing $CsPbBr_3$ (ASC18) NC solid samples. Spectroscopic characterization at MIT was supported by the U.S. Department of Energy, Office of Science, Basic Energy Sciences, under award no. DE-SC0019345. Thermal transport modeling was supported by the U.S. National Science Foundation under award 1452857. N.N.W. was partially supported by a MathWorks Engineering Fellowship. The opinions and views expressed in this publication are from the authors and not necessarily from MathWorks. E.R.P. was supported by the US Department of Defense through the National Defense Science & Engineering Graduate (NDSEG) Fellowship Program. S. S. acknowledges the funding support of the Swiss National Science Foundation under projects PYAPP2_166897/1 and P500PN_202653. A.P.W. acknowledge support for this research from the U.S. Department of Energy (DOE), Office of Science, Basic Energy Sciences under Award DE-SC0019998.

# Persistent Enhancement of Exciton Diffusivity in CsPbBr$_3$ Nanocrystal Solids

Wenbi Shcherbakov-Wu[1], Seryio Saris[2,†], Thomas Sheehan[3], Narumi Nagaya Wong[3], Eric R. Powers[3], Franziska Krieg[4,5,††], Maksym V. Kovalenko[4,5], Adam P. Willard,[1] William A. Tisdale[*,3]

1. Department of Chemistry, Massachusetts Institute of Technology, Cambridge, MA, United States
2. Laboratory of Nanochemistry for Energy (LNCE), Institute of Chemical Sciences and Engineering (ISIC), École Polytechnique Fédérale de Lausanne, CH-1950 Sion, Switzerland
3. Department of Chemical Engineering, Massachusetts Institute of Technology, Cambridge, MA, United States
4. Department of Chemistry and Applied Bioscience, ETH Zürich, Zürich, Switzerland
5. Laboratory for Thin Films and Photovoltaics and Laboratory for Transport at Nanoscale Interfaces, Empa – Swiss Federal Laboratories for Materials Science and Technology, Dübendorf, Switzerland

*Correspondence: tisdale@mit.edu

†Current affiliation: Chair in Hybrid Nanosystems, Faculty of Physics, LMU Munich, Munich, Germany

††Current address: Avantama AG, Laubisrütistr. 50, 8712 Stafa Switzerland,

## Supplementary Note 1: Material synthesis and characterization
### $CsPbBr_3$ NCs (OLA-OA) and $CsPbBr_3$ NCs (ALOx):

**Chemicals:** Cesium carbonate ($Cs_2CO_3$, 99.9%, Sigma Aldrich), Lead (II) bromide ($PbBr_2$, 99.9%, Alfa Aesar), oleylamine (OLAM, tech.grade 70%, Sigma Aldrich), oleic acid (OLAC, 90% technical grade, Sigma Aldrich), Trimethylaluminium (TMA, 98%, Strem), ethanol (EtOH, anhydrous, 95%, Sigma Aldrich), methanol (95%, Sigma Aldrich), Octadecene (ODE, 90% technical grade, Acros), Octane (anhydrous, > 99%, Sigma Aldrich), Hexane (anhydrous, >96%, TCI), Acetone (anhydrous extra dry, syn. Grade, Acros Organics).

**Perovskite Nanocrystal Synthesis:** $CsPbBr_3$ NCs were colloidally synthesized by a hot-injection method following previously reported and detailed procedures.[1-2] The resulting NCs were washed in two-steps; (1) centrifugation at 6000 rpm (~ 20 mins) and redispersion in anhydrous hexane, and (2) addition of anhydrous acetone (0.5 volume ratio with hexane), centrifugation at 6000 rpm (~ 5 mins) and redispersion in equal volumes of anhydrous hexane and octane, or just octane. Further dilutions were performed in accordance with the film preparation process.

**Colloidal Atomic Layer Deposition (c-ALD) on Perovskites NCs:** Colloidal perovskite NCs were covered with amorphous aluminum oxide shells of different thicknesses following the previously reported and optimized c-ALD method developed by Loiudice *et al.*[3-4] An octane solution of perovskite NCs (typically with a NC concentration of ~8 mM) was placed in a three-necked flask connected to a standard $N_2$ Schlenk line. The c-ALD surface treatment was then performed as follows: (1) the dropwise addition of 1 mL of TMA in octane (1.6 mM) to the reaction flask with a speed of 1 mL/h, (2) a 5 min waiting time to ensure that the reaction in step 1 was completed, (3) addition of $O_2$ gas by means of a mass flow controller. This three-step process is referred to as a single c-ALD cycle. The cycle is then repeated 4, 8 and 12 times to achieve increasing aluminum oxide shell thickness, which was then measured through dynamic light scattering.

**Perovskite Nanocrystal Film Preparation:** $CsPbBr_3$ NCs, both with and without c-ALD surface treatment, were deposited on 15 x 15 x 0.5 mm quartz substrates via spin-coating. All steps were performed in the glovebox under inert $N_2$ environment. Prior to spin-coating, the substrates were sonicated with consecutive cycles of acetone/isopropanol, and then treated with a toluene solution of (3-mercaptopropyl)trimethoxysilane (0.02 M) for 12h to improve NC adhesion to the surface. $CsPbBr_3$ NC solutions with hexane:octane (1:1 by volume) solvent ratio were spin-coated at 1000 rpm for 45s to achieve optical densities of 0.01-0.07 at the first excitonic transition.

**Atomic Layer Deposition (ALD):** Low temperature ALD was performed on a Savannah-200 ALD system from Cambridge Nanotech Inc., following the method from published work.[5] Briefly, amorphous AlOx was deposited on top of the $CsPbBr_3$ NC films. Trimethylaluminum (TMA) and ultrapure water were used as aluminum and oxygen sources, respectively. The reaction chamber was kept at a temperature of 50 °C and an operating pressure of ~0.10 Torr. The thickness of the deposited ALD layer was controlled by varying the number of ALD cycles, with 100 cycles equating to ~10 nm overcoating layer on top of the $CsPbBr_3$ NC film.

**Atomic Force Microscopy (AFM):** AFM was used to measure the overall thickness and roughness of the perovskite composite films. The measurements were performed using a Nanoscope IIIa (Veeco, USA), operated in tapping-mode, with Nanosensors PP-NCSTR AFM probes. Thin lines were scratched on the samples to reveal the Si substrate. The mapping was carried-out at the edge of the lines.

**Transmission Electron Microscopy (TEM):** TEM images were acquired on an Analytical JEOL-2100F FETEM equipped with a Gatan camera, using a beam energy of 120 kV. NC samples were dropcasted on Cu TEM grids (Ted Pella Inc.) prior to the imaging. Size measurements were performed using the software Image J and counting 200 particles per sample.

**Steady-State Absorption:** Steady-state UV−Visible absorption measurements were performed in transmission mode using a PerkinElmer Lambda 950 spectrophotometer equipped with deuterium and tungsten halide lamps for UV and Vis-IR ranges, respectively. A PMT and Peltier-controlled PbS were used for detection.

**Steady-State Photoluminescence (PL) Spectroscopy:** Steady-state emission and quantum yield (QY) PL measurements were recorded by a Horiba Jobin Yvon Fluorolog-3 spectrometer equipped with a PMT detector. All PL spectra were collected at an excitation wavelength of 370 nm. Absolute QY measurements were performed in a Spectralon® coated integrating sphere. For each sample, four measurements were performed: (i) sample emission ($S_{em}$); (ii) blank glass emission ($B_{em}$) (iii) sample excitation ($S_{exc}$) and (iv) blank glass excitation ($B_{exc}$). The absolute QY was then calculated as follows:

$$QY = \frac{S_{em} - B_{em}}{B_{exc} - S_{exc}} \qquad (1)$$

The reported QY values is the average of three measurements.

*CsPbBr$_3$ NCs (zwitterionic ligands):*
**Chemicals:** The following reagents were used as received. Cesium carbonate $Cs_2CO_3$ from Fluorochem, 3-(N,N-dimethyldodecylammonio)propanesulfonate (ASC12, >99%) from Roth, lead acetate trihydrate (99.99%), bromine (99.9%), 1-octadecene (ODE, technical grade), 3-(N,N-dimethyloctadecylammonio)propanesulfonate (>99%, ASC18) and oleic acid (90%, OA) from Sigma Aldrich/Merck, toluene (for synthesis), acetone (HPLC grade), ethylacetate (HPLC grade) from Fischer and trioctylphosphine (>97%, TOP) from STREM.
**Cs-oleate 0.4 M in ODE:** Cesium carbonate (1.628 g, 5 mmol), oleic acid (5 mL, 16 mmol) and 1-octadecene (20 mL) were evacuated at 25-120 °C until the completion of gas evolution.
**Pb-oleate 0.5 M in ODE:** Lead (II) acetate trihydrate (4.607g, 12 mmol), oleic acid (7.6 mL, 24 mmol) and 1-octadecene (16.4 mL) were mixed in a three-necked flask and evacuated at 25-120 °C until the complete evaporation of acetic acid and water.
**TOP-Br2 0.5 M in toluene:** TOP (6mL, 13 mmol) and bromine (0.6 mL, 11.5 mmol) were mixed under inert atmosphere. Once the reaction was complete and cooled to room temperature, the TOP-Br2 was dissolved in toluene (18.7 mL).
**CsPbBr3 Nanocrystals with ASC18 as a ligand**: CsPbBr$_3$ NCs were synthesized by dissolving Cs-oleate (4 mL, 1.6 mmol), Pb-oleate (5 mL, 2.5 mmol) and ASC18 (0.215 g, *ca.* 0.512 mmol) in ODE (5 mL) and heating the mixture under vacuum to 130 °C, whereupon the atmosphere was changed to argon and TOP-Br$_2$ in toluene (5 mL, 5 mmol of Br) was injected. The reaction was cooled immediately by an ice bath.

The crude solution (19 mL) was precipitated by the addition of acetone (10 mL) in a nitrogen filled glovebox, followed by the centrifugation at 29500g (g is the earth gravity) for 10 minutes. The precipitated fraction was dispersed in toluene (3 mL) and then washed three more times. Each time the solution was mixed with two volumetric equivalents of acetone and centrifuged at 1300 g for 10 minutes, and subsequently dispersed in progressively smaller amounts of the solvent (1.5mL for the second cycle, 0.75 mL for the third cycle). After the last precipitation, NCs were dispersed in 1 mL of toluene and centrifuged at 1300 g for 1 minute to remove any non-dispersed residue.

**Film formation:** Films were spin-coated on single crystalline quartz substrates that were previously cleaned. Specifically, they were sonicated in soap water, cleaned with water stream, and blow dried. This sequence was repeated twice. The films were then sonicated in ethanol, blow dried, sonicated in acetone and blow dried. The films were subsequently covered with a monolayer of HMDS and annealed in a nitrogen filled glovebox. 30 uL of the solution were used per cm$^2$ of substrate, they were deposited, first spun at 500 rpm for 10s followed by 1 min at 2000 rpm. For spincoating the NC solutions were diluted to 3 mg/mL inorganic NC mass.

**Instrumentation:** UV-Vis absorption spectra of colloidal NCs were collected using Jasco V670 and Jasco V770 spectrometers in transmission mode. Fluorolog iHR 320 Horiba Jobin Yvon was used to acquire steady-state PL spectra from solutions, using excitation at 350nm. PL QYs of films and solutions were measured with Quantaurus-QY Absolute PL quantum yield spectrometer from Hamamatsu. TEM images were collected using Hitachi HT7700 microscope operated at 100 kV.

*Steady-state PL Spectroscopy:*
The steady-state PL spectroscopy was conducted with the same excitation conditions and sample mounting configuration as the TPLM measurements. After the tube lens, the emission is diverted using a silver mirror, collimated, refocused, coupled into a multimode fiber (Thorlabs, BFL 105LS02), and dispersed by a 300 gr/mm grating inside a Princeton Instrument SP2150 spectrograph. Spectra were recorded using a thermoelectrically-cooled CCD camera (Princeton Instrument Pixis 100).

**Supplementary Note 2: Theoretical Model and Kinetic Monte Carlo Simulations:**
*Description of the model:*
The NC solid was modeled as a square lattice with each lattice site representing an individual NC. Simulations were carried out on a 1000x1000 lattice to avoid any boundary affects. The lattice spacing was set to $l = 10$nm. Each NC adopts one of two structural configurations: a ground configuration (gnd) and a rapid modified configuration (mod). Nanocrystals in the mod state stochastically transition to the ground state with a rate $k_{\text{relax}} = \tau_{\text{relax}}^{-1}$, where $\tau_{\text{relax}} = 8.1\mu s$. Excitons occupy individual NC and diffuse through nearest-neighbor hopping, as described below. Excitons decay stochastically with a rate $k_{\text{decay}} = \tau_{\text{ex}}^{-1}$, where $\tau_{\text{ex}} = 2.5$ns is the exciton lifetime.

Excitons are created in pulses at increments of $\tau_{\text{pulse}}$, which is an adjustable parameter that sets the pulse frequency. At each increment, excitons are created at every lattice site in the system with Gaussian spatial distribution,
$$P_{\text{excite}}(x,y) = \alpha_{\text{pulse}} \exp(-r_{\text{center}}^2 / 2\sigma_{\text{pulse}}^2),$$
where $x$ and $y$ denote the coordinates of a given lattice site (in units of $l$), $\alpha_{\text{pulse}}$ represents the intensity of the laser pulse ($0 < \alpha_{\text{pulse}} < 1$), $r_{\text{center}}^2 = (x - 500)^2 + (y - 500)^2$, and $\sigma_{\text{pulse}} = 10.62$, to yield a full-width half-max of 250nm.

### *Model dynamics:*

Model dynamics were simulated with a simple Monte Carlo algorithm. Initially, there are no excitons and every NC is in the gnd configuration. Whenever an exciton occupies a NC, via laser pulse or hopping, the NC immediately transitions into the mod configuration. Exciton hopping dynamics are then simulated with a 1ps timestep until all excitons have decayed. At each timestep, each exciton attempts a hop to a randomly selected neighbor and hops with probability $P_{hop}^{(mod)} = 0.7$ or $P_{hop}^{(gnd)} = 6 \times 10^{-3}$ if the neighbor is in the modified or ground configuration, respectively. Following this, each exciton randomly decays (is eliminated from the simulation) with probability $P_{decay} = 4 \times 10^{-4}$ (selected to yield an exciton lifetime of 2.5ns). Excitons are constrained to single occupancy, so if an exciton attempts to hop to an occupied site, its hopping probability is zero.

After all excitons have decayed, the NC configurations are relaxed over the time increment to the next pulse. To do this we select a random relaxation time, $\Delta t_{relax}$, for each NC in the mod configuration. Selecting $\Delta t_{relax} = -\tau_{relax} \ln(1 - f)$, where $f$ is a random number between 0 and 1 yields a proper exponential distribution of configurational lifetimes. For a given NC, if $\Delta t_{relax} \leq \tau_{pulse}$, then the NC is returned to the ground state for the next simulated laser pulse. If $\Delta t_{relax} > \tau_{pulse}$, the NC remains in the mod state. Thus, as the pulse frequency nears $\tau_{relax}$, excitons can hop to NCs that transitioned into a mod state during the previous pulse.

For each set of $\alpha_{pulse}$ and $\tau_{pulse}$, we simulated a sequence of 1000 laser pulses, recording exciton positions every 50 timesteps. Time-dependent exciton density, $\rho_{ex}(x, y, t)$ was computed by averaging over the occupancy state of site $(x, y)$ at time $t$ after the most recent laser pulse for all 1000 pulses. Exciton diffusivity was computed based on the shape of this distribution, as described in the main text.

**Supplementary Note 3: Transient Photoluminescence Microscopy Method and Discussion**

The nanocrystal solid samples were mounted on a piezo stage (attocube, ANC350) in a closed-cycle liquid helium cryostat under vacuum (Cryostation, Montana Instruments). For low-temperature measurements, the samples were cooled down to 5 K before being heated up with a temperature controller (Lakeshore Model 335) to designated temperatures. Transient PL microscopy was performed using a home-built fluorescence microscope, as shown in **Fig. S1**. NC solids were excited with a 405 nm pulsed laser (PDL 800-D, pulse width < 100 ps) at various repetition rates and fluences. The excitation laser pulses were spatially filtered by a single-mode optical fiber and focused down to a near diffraction-limited spot by an objective lens mounted inside the vacuum chamber (Zeiss EC Epiplan-Neofluar 100X/0.85 NA). Epifluorescence was collected by the same objective and filtered by a dichroic mirror (Semrock, Di02-R405) and a longpass colored glass filter (Thorlabs, FGL435M). The emission was then passed through a tube lens (Thorlabs, TTL200-S8) and a telescope (Thorlabs, AC254-030-A and AC254-125-A). The Avalanche Photo Diode detector (APD, Micro Photon Devices, timing resolution ~50 ps, active area 50 μm × 50 μm) was positioned at the imaging plane (495x magnified) after the telescope. The position of the APD was controlled by two orthogonal motorized actuators (Thorlabs, ZFS25B). The evolution of the photoluminescence spatial profile with time was acquired by scanning the detector across the magnified emission profile and collecting a photoluminescence decay histogram at each position. For the samples in this study, a single dataset could be collected within 15~20 minutes.

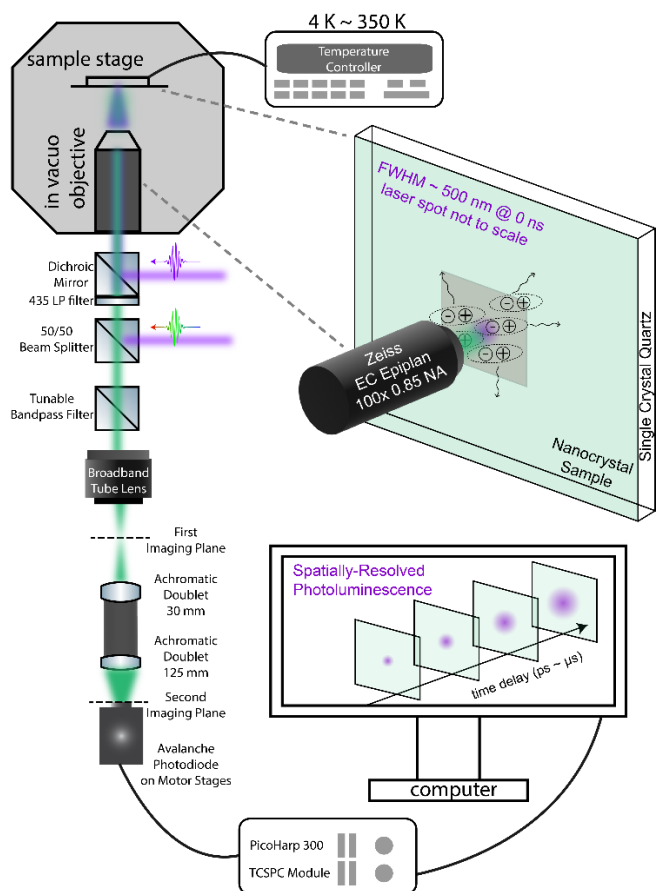

**Fig. S1.** Schematic of the home-build cryomicroscope.

The diffusion data were taken as a function of time delay and spatial position. The general diffusion equation in one dimension is as follows:

$$\frac{\partial n(x,t)}{\partial t} = D(t)\frac{\partial^2 n(x,t)}{\partial x^2} - k(t)n(x,t) \tag{1}$$

where $n(x,t)$ is the exciton density distribution as a function of time and location, $D(t)$ is the time-dependent exciton diffusivity, and $k(t)$ is the time-dependent exciton decay rate. For CsPbBr$_3$ NCs, we assume that the diffusivity and decay rate are time-independent during the first 3 ns. Additionally, we treat the laser pulse as an instantaneous source, $n(\tilde{x},0)$. Then, the general solution to Eqn. (2) could be written as:

$$n(x,t) = e^{-kt}\frac{1}{\sqrt{4\pi Dt}}\int_{-\infty}^{\infty} n(\tilde{x},0)\, e^{-\frac{(x-\tilde{x})^2}{4Dt}}\, d\tilde{x} \tag{2}$$

As we normalize the PL emission at any given time delay, Eqn. (3) can be re-written into Eqn. (3) as the following:

$$n(x,t) \propto \int_{-\infty}^{\infty} n(\tilde{x},0)\frac{1}{\sqrt{4\pi Dt}}e^{-\frac{(x-\tilde{x})^2}{4Dt}}d\tilde{x} = n(\tilde{x},0) * G(x,t) \tag{3}$$

where $G(x,t) = \frac{1}{\sqrt{4\pi Dt}}e^{-\frac{x^2}{4Dt}}$. Eqn. (3) shows that at any given time, the exciton density distribution is the convolution of the initial exciton density distribution created by the laser pulse and the Gaussian function. The variance of the Gaussian $G(x,t)$ is $\sigma^2(t) = 2Dt$. Therefore, by fitting the spatial profile at each time delay (**Fig. S2**), the diffusivity could be extracted as:

$$D = \frac{\sigma^2(t) - \sigma_0^2(t)}{2t} \tag{4}$$

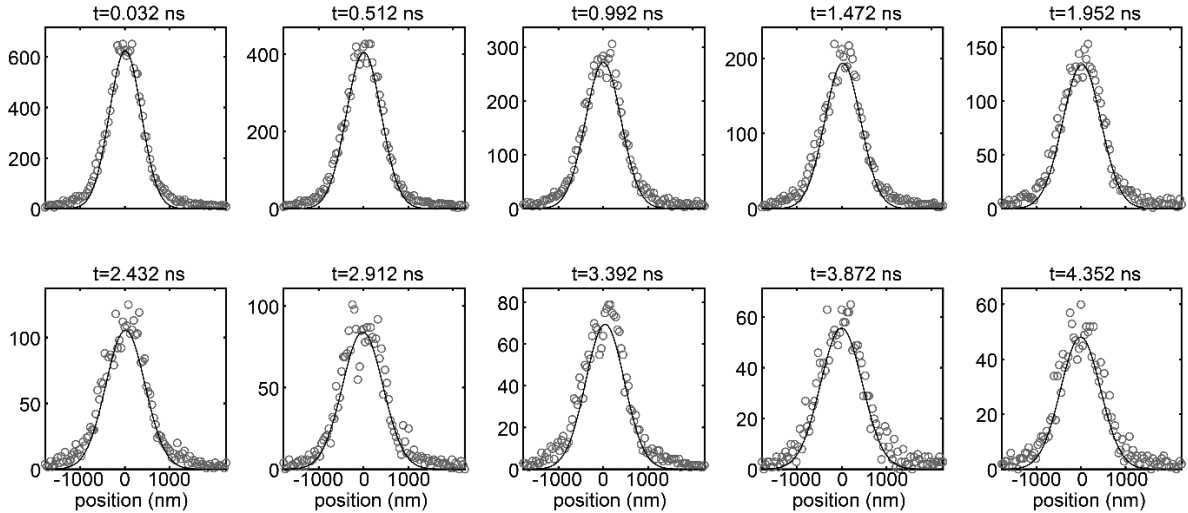

**Fig. S2.** Sample time slices of the TPLM measurements in CsPbBr$_3$ NC solids at 25 nW. Open circles are data points and solid lines are the fitted Gaussian shapes.

In the MSD plots, the error bars on individual data points correspond to the standard deviations of the 25 neighboring data points. A line intercepting (0,0) was fitted to the data points between 0 and 3 ns, by which time more than 80% of the excitons have recombined. The error bars on diffusivity are from the linear fitting.

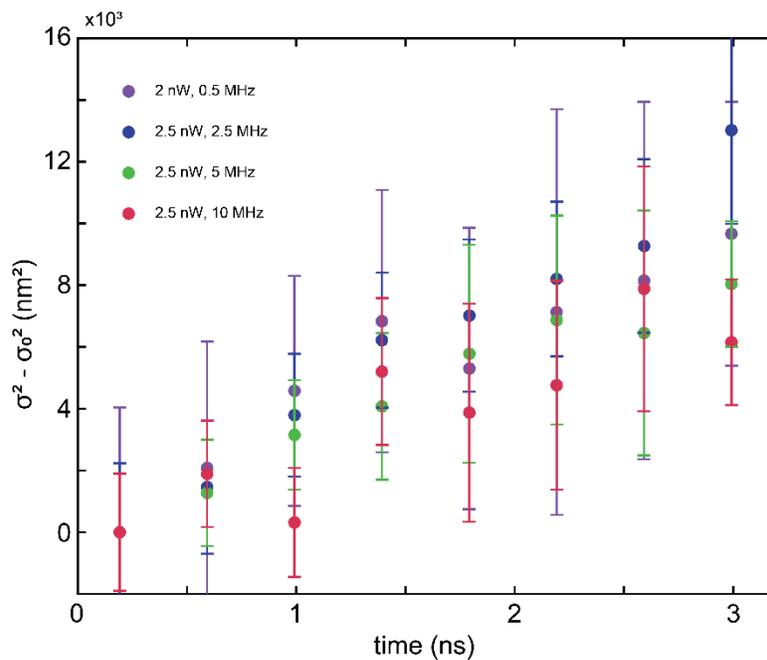

**Fig. S3.** Mean square displacement (MSD) curves while holding the time-averaged power relatively constant and changing the repetition rates and fluences. The TPLM data are shown in **Fig. 2e**.

**Supplementary Note 4: Sample thickness**

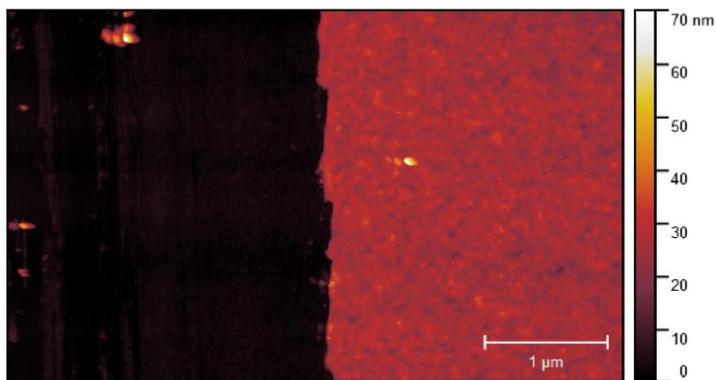

**Fig. S4.** Atomic Force Microscopy (AFM) image of sample NC solids (OLA-OA). The NC film has a thickness of 30~40 nm, and is relatively smooth and uniform.

# Supplementary Note 5: Spatially-integrated steady-state and transient photoluminescence characterization

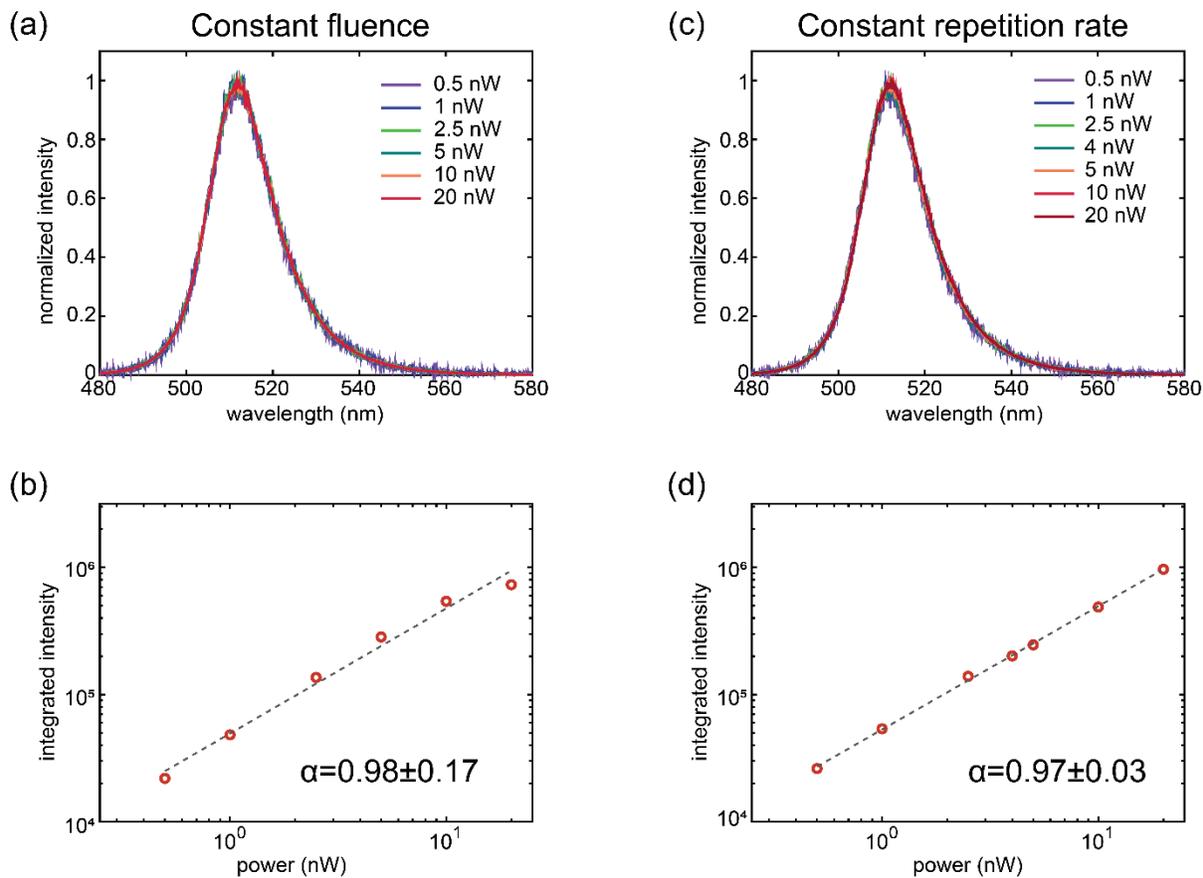

**Fig. S5.** Steady-state photoluminescence spectra of the $CsPbBr_3$ NC solid film at various power conditions. (a) Normalized PL spectra under constant fluence but various repetition rates. (b) Integrated intensity as a function of power (constant fluence, various repetition rates). The dash line is a linear fit to the data points with the exponent shown in the panel. (c) Normalized PL spectra under constant repetition rate but various fluences. (d) Integrated intensity as a function of power (constant repetition rate, various fluences). The dash line is a linear fit to the data points with the exponent shown in the panel.

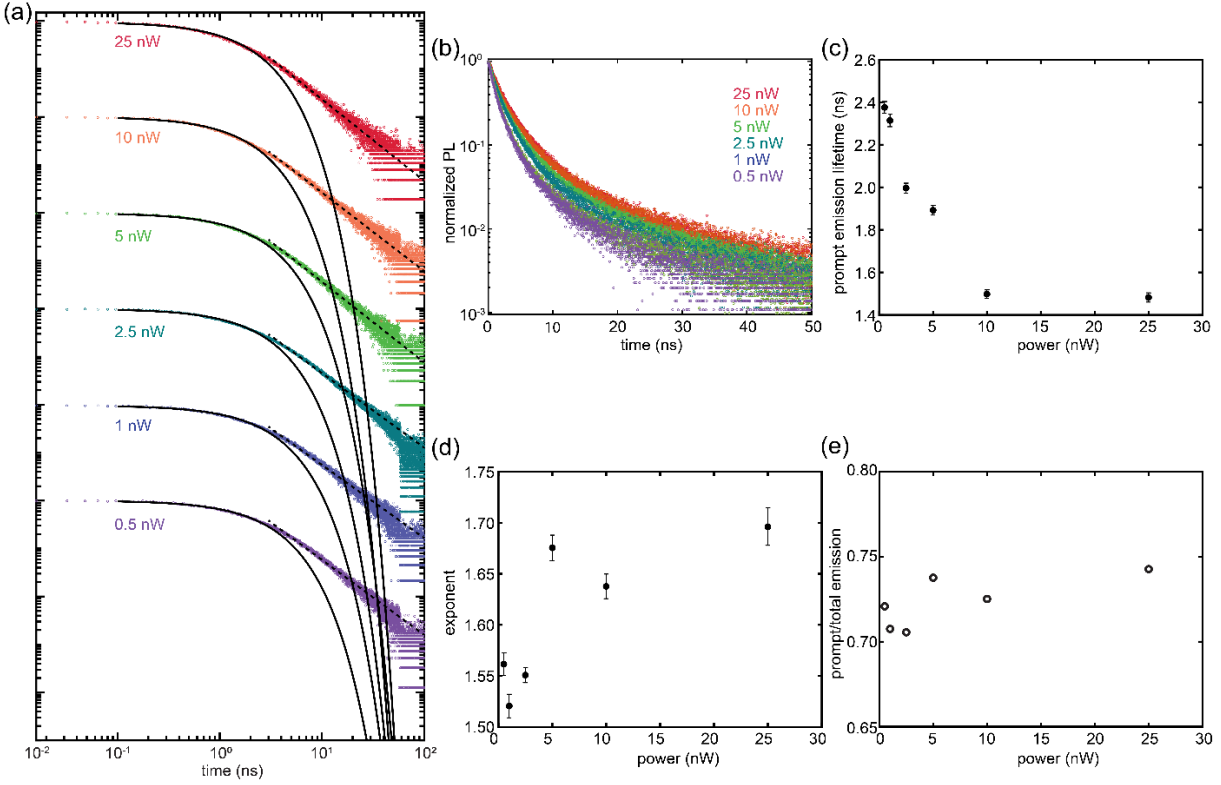

**Fig. S6.** Transient photoluminescence traces at various time-averaged power. (a-b) Power-dependent transient PL plotted on (a) log-log or (b) semi-log scale (all taken at 5 MHz). The solid lines are single exponential fits to the first 3 ns, and dash lines are power law fits to 3~20 ns. Same as Fig. 1d. (c-d) Extracted lifetime from the single exponential fit and extracted exponent from the power law fit for the two different time regimes. (e) The ratio between the prompt emission and total emission at various power.

## Supplementary Note 6: Calculation of excitation density

All TPLM measurements were conducted in a home-built microscope (see Methods). The focused laser spot on the sample surface is approximately Gaussian in spatial intensity profile with a 500 nm diameter (full width at half maximum). Nominal powers referenced here and in the main text were measured outside the microscope. The laser power throughput of the microscope was measured to be 37.4%, meaning that if nominal power is $1\ nW$, then the power at sample surface is

$$1\ nW * 37.4\% = 0.37\ nW.$$

The intensity of the laser power at the sample surface is then

$$\frac{0.37\ nW}{250 nm^2 * \pi} = 0.19\ W/cm^2.$$

Reported values of laser intensity, fluence and excitation density correspond to the true value at sample surface. For a repetition rate of 2.5 MHz and nominal laser power of 1 nW, the laser pulse fluence is

$$\frac{0.19 W/cm^2}{2.5*10^6\ s^{-1}} * \frac{\mu W}{10^{-6} W} = 0.076 \mu J/cm^2.$$

To calculate excitation density the absorption coefficient, as well as the absorption cross section, must be determined first. For CsPbBr$_3$ NCs with an 8.3 nm edge length, the absorption coefficient at 400 nm (in toluene) was determined from the size-dependent molar extinction coefficient[6],

$$0.0242 cm^{-1}\mu M^{-1} nm^{-3} * (8.3\ nm)^3 * \frac{10^6 \mu M}{1 M} = 1.38 * 10^7 cm^{-1} M^{-1}.$$

The absorption cross section of a single 8.3 nm CsPbBr$_3$ NC at 400 nm is then,

$$1.38 * 10^7 cm^{-1} M^{-1} * \frac{ln(10)*10^3 cm^3 L^{-1}}{6.02*10^{23} mol^{-1}} = 5.29 * 10^{-14} cm^2.$$

We assume that the absorption cross section at 405 nm is the same as this value, within error.

Then, the average number of photons absorbed per NC per laser pulse, <N>, (also referred to in the text as a unitless "excitation density") is,

$$\langle N \rangle = photons\ absorbed\ per\ NC\ per\ pulse$$

$$= absorption\ cross\ section\ [cm^2]$$
$$* incident\ photons\ per\ laser\ pulse\ per\ unit\ area\ [cm^{-2}]$$

$$= 5.29 * 10^{-14} cm^2 * \frac{0.076(\mu J cm^{-2}) * \frac{1 J}{10^6 \mu J}}{\left(\frac{1240 eV * nm}{405\ nm}\right) * \frac{1.602 * 10^{-19} J}{1 eV}}$$

$$= 0.0082$$

**Table S1.** Experimental conditions for the power-dependent photoluminescence spectra shown in Fig. 1c: excitation at 405 nm, 2.5 MHz, spot size FWHM ~500 nm.

| Nominal power (nW) | Actual laser power @ sample surface (nW) | Intensity @ sample surface (W/cm$^2$) | Fluence @ sample surface (μJ/cm$^2$) | $<N>$ = Excitation density @ sample surface (exciton generated per NC per pulse) |
|---|---|---|---|---|
| 0.5 | 0.185 | 0.095 | 0.038 | 0.0041 |
| 1 | 0.37 | 0.19 | 0.076 | 0.0082 |
| 2.5 | 0.925 | 0.475 | 0.19 | 0.0205 |
| 4 | 1.48 | 0.76 | 0.30 | 0.0326 |
| 5 | 1.85 | 0.95 | 0.38 | 0.041 |
| 10 | 3.7 | 1.9 | 0.76 | 0.082 |
| 20 | 7.4 | 3.8 | 1.52 | 0.163 |

**Table S2.** Experimental conditions for the transient photoluminescence data shown in Fig. 1d: excitation at 405 nm, 5 MHz, spot size FWHM ~500 nm.

| Nominal power (nW) | Actual laser power @ sample surface (nW) | Intensity @ sample surface (W/cm$^2$) | Fluence @ sample surface (μJ/cm$^2$) | $<N>$ = Excitation density @ sample surface (exciton generated per NC per pulse) |
|---|---|---|---|---|
| 0.5 | 0.185 | 0.095 | 0.019 | 0.00205 |
| 1 | 0.37 | 0.19 | 0.038 | 0.0041 |
| 2.5 | 0.925 | 0.475 | 0.095 | 0.01025 |
| 5 | 1.85 | 0.95 | 0.19 | 0.0205 |
| 10 | 3.7 | 1.9 | 0.38 | 0.041 |
| 25 | 9.25 | 4.75 | 0.95 | 0.1025 |

## Supplementary Note 7: Data simulation

As the diffusivities measured in these samples were low, we turned to data simulation to confirm the validity of the analyses (**Fig. S7**). We assumed an initial spot size of 600 nm in FWHM, full-width half-max, ($\sigma(t=0) = \frac{600\ nm}{2*\sqrt{2*\ln(2)}}$) and an exciton diffusivity of 0.01 cm²/s, similar to the experimental conditions. An experimentally obtained normalized lifetime trace $I(t)$ was added to simulate the data decay as a function of time. Therefore, the spatial profile at the laser arrival time could be expressed as:

$$n(x, t=0) = \frac{A * I(t=0)}{\sqrt{2\pi} * \sigma(t=0)} * e^{-\frac{x^2}{2\sigma(t=0)^2}} \quad (8)$$

where $A$ is a pre-factor modulating the intensity of the signal. Here, we arbitrarily set $A = 2 * 10^5$ to match the experimental signal intensity. At later times $t$, the variance and spatial distribution could be written as:

$$\sigma(t) = \sqrt{\sigma(t=0)^2 + 2Dt} \quad (9)$$

$$n(x, t) = \frac{A * I(t)}{\sqrt{2\pi} * \sigma(t)} * e^{-\frac{x^2}{2\sigma(t)^2}} \quad (10)$$

Both ambient background noise and shot noise were considered and added to the simulated PL emission data (**Fig. S7a**). Specifically, ambient background noise refers to both the dark current of the detector and the photons from the experiment surroundings that reached the detector. To recreate the noise, we created a matrix of uniformly distributed random numbers with the intensity similar to experimentally observed dark counts. In addition, shot noise describes the fluctuation of number of photons detected as photon detections are individual events. Shot noise is proportional to the square root of the experimental signal as:

$$I_{shot} \propto \sqrt{signal} \quad (11)$$

For each data point, the shot noise is recreated as:

$$I_{shot} \propto rand(0,1) * \sqrt{signal} \quad (12)$$

where $rand(0,1)$ is a random number generated between 0 and 1.

The same data analysis and fitting procedure were then applied to the simulated data set, and a diffusivity of 0.01 cm²/s was recovered (**Fig. 7b-c**). It should be noted that only when the full spatial profile (when there were negligible PL counts towards both ends) was fitted and when the background counts were properly accounted for, could the diffusivity be accurately recovered; otherwise, a value significantly larger than 0.01 cm²/s was extracted.

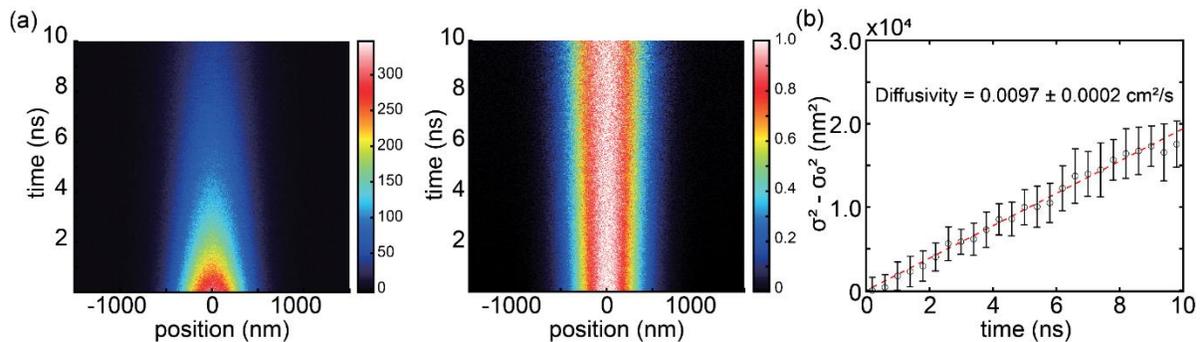

**Fig. S7.** Simulated data and analysis. (a) 2D color maps of unnormalized (left) and normalized (right) simulated transient photoluminescence microscopy data considering both dark counts and shot noise, assuming exciton diffusivity $D = 0.01\ cm^2/s$ and initial spot size (FWHM) of 600 nm. (b) Extracted mean square displacement as a function of time, recovering a diffusivity nearly identical to the simulation input.

## Supplementary Note 8: Heat transport simulation

To examine the effect of laser-induced heating on the nanocrystal temperature under experimental conditions, the temperature as a function of time and position in the nanocrystal was simulated. Using cylindrical coordinates, the time and spatial evolution of heat in the nanocrystal was modeled via the following partial differential equation:

$$\frac{\partial \Theta(r,z,t)}{\partial t} = D_{thermal} \nabla^2 \Theta(r,z,t) \tag{5}$$

$\Theta(r,z,t)$ is the dimensionless temperature, where $\Theta = \frac{T(r,t) - T_{initial}}{\Delta T_{max}}$ as a function of radial position $r$, axial position $z$ and time $t$. $T$ is the actual temperature in the sample, $T_{initial}$ is the initial temperature that the sample is at before the laser excitation and $\Delta T_{max}$ is the maximum temperature rise induced in the sample as a result of the laser excitation. $D_{thermal}$ is the thermal diffusivity of the nanocrystal array; the representative value was obtained from Yang *et al.*[7]

To calculate the most extreme scenario, it was assumed that all of the energy from the laser pulse was converted to heat energy instantaneously in the sample. Additionally, no-flux boundary conditions were used for all edges of the sample. The initial condition was derived from the measured experimental data by fitting a Gaussian shape to match the radial distribution of heat, and the axial heat distribution was approximated using Beer's law absorption at the excitation wavelength.

The partial differential equation was solved using an explicit finite differencing method. The results showed that the time required for the sample to reach less than 5% of the initial dimensionless temperature was on the same order of magnitude as the inverse of the repetition rate. The maximum temperature rise induced in the sample as a result of the laser excitation $\Delta T_{max}$ was calculated using the experimental conditions:

$$\Delta E_{photon} \times \langle n \rangle = \rho_{bulk} c_{p,bulk} V_{NC\,bulk} \Delta T_{max} \tag{7}$$

$\Delta E_{photon}$ is the photon energy, $\langle n \rangle$ is the average number of photons absorbed per pulse per nanocrystal, $\rho_{bulk}$ is the bulk density of the nanocrystal, $c_{p,bulk}$ is the bulk specific heat capacity of the nanocrystal, $V_{NC\,bulk}$ is the bulk volume of the nanocrystal (using experimental dimensions). The resulting $\Delta T_{max}$ was found to be on the order of 0.01K, which is vanishingly small.

Despite the time required for the sample to return back to less than 5% of its initial dimensionless temperature being on the same order of magnitude as the inverse of the repetition rate, the effect of laser-induced heating on the temperature of the sample is expected to be negligible because the actual temperature rise induced in the sample is very low due to the low laser fluences used.

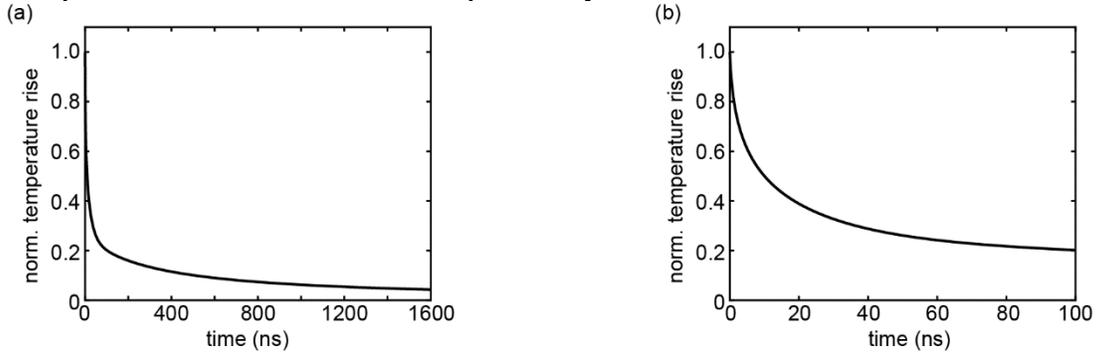

**Fig. S8.** Simulated heat transport as a result of laser heating on the sample. (b) shows the same data as panel (a), with a zoom-in view of the early-time dynamics.

**Supplementary Note 9: Characterization of samples with various surface treatments**

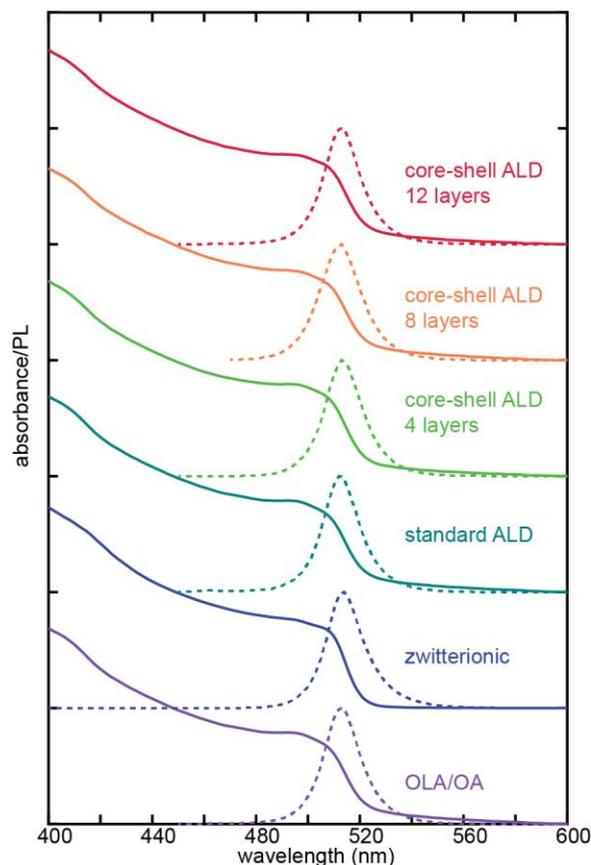

**Fig. S9.** Steady-state optical characterization of all the CsPbBr$_3$ NC solids studied. The solid lines correspond to absorbance and dash lines correspond to photoluminescence.

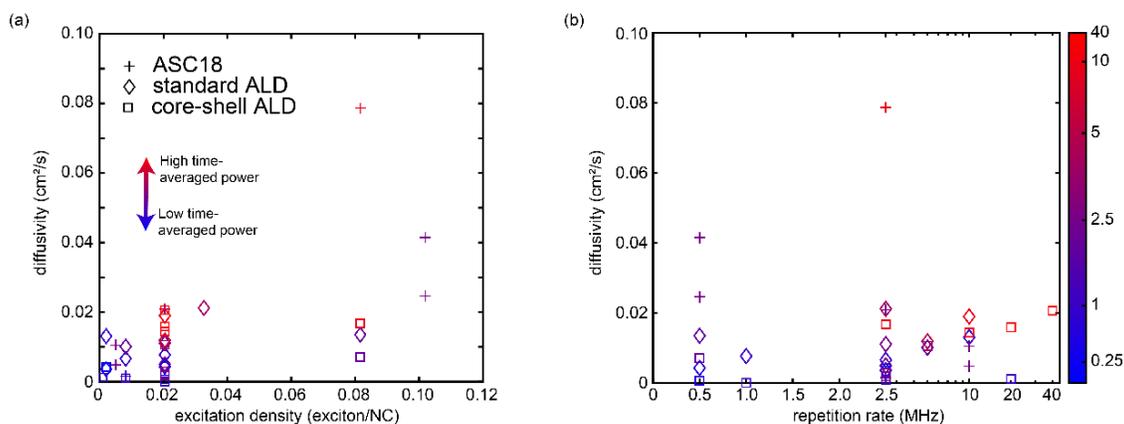

**Fig. S10.** Exciton diffusivity of samples with different surface treatments denoted with different markers. (a) Exciton diffusivity as a function of fluence. (b) Exciton diffusivity as a function of laser repetition rate. Data point colors in (a) and (b) correspond to the time-average power. The x-axis is in linear scale when repetition rate is less than 2.5 MHz and logarithm scale when repetition rate is above 2.5 MHz. Color map is shown on the right.

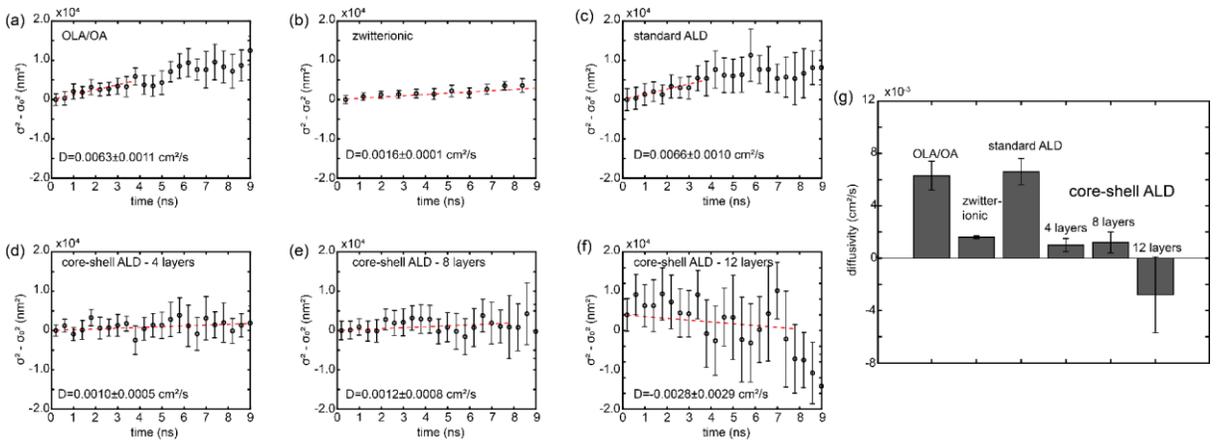

**Fig. S11.** Mean square displacement as a function of time for all CsPbBr$_3$ NC solid samples under 1 nW. (a) OLA/OA capped CsPbBr$_3$ NCs. (b) Zwitterionic ligand (ASC18) capped CsPbBr$_3$ NCs. (c) CsPbBr$_3$ NCs with standard AlOx ALD layer. (d) CsPbBr$_3$ with colloidal AlOx ALD layers (4 layers). (e) CsPbBr$_3$ with colloidal AlOx ALD layers (8 layers). (f) CsPbBr$_3$ with colloidal AlOx ALD layers (12 layers). (g) Bar graph showing diffusivities of different samples under ~ 1 nW.

## Supplementary Note 10: Diffusivity relaxation curve on the oxide-coated CsPbBr$_3$ NCs

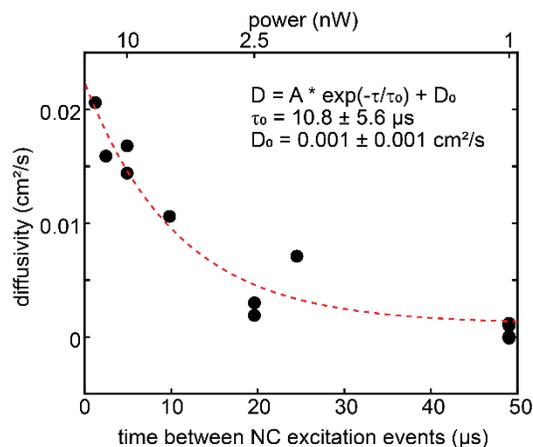

**Fig. S12.** Relaxation curve of colloidal AlOx-coated CsPbBr$_3$ NCs. Exciton diffusivity of CsPbBr$_3$ NCs with 4 layers of colloidal AlOx ALD coating as a function of time between NC excitation events. Mirrored x-axis shows the corresponding time-averaged power. The red dash line is a single exponential decay fit. The fitted equation and parameters are indicated within the plot.